\documentclass[]{spie}  

\usepackage{amsmath,amsfonts,amssymb}
\usepackage{graphicx}
\usepackage[colorlinks=true, allcolors=blue]{hyperref}

\title{Reconstruction methods for the phase-shifted Zernike wavefront sensor}

\author[a]{Vincent Chambouleyron}
\author[b]{Mahawa Cissé}
\author[a]{Maïssa Salama}
\author[c,d]{Sebastiaan Haffert}
\author[e]{Vincent Déo}
\author[f]{Charlotte Guthery}
\author[g]{J. Kent Wallace}
\author[a]{Daren Dillon}
\author[a]{Rebecca Jensen-Clem}
\author[a]{Phil Hinz}
\author[a]{Bruce Macintosh}

\affil[a]{University of California Santa Cruz, 1156 High St, Santa Cruz, USA}
\affil[b]{Aix Marseille Univ, CNRS, CNES, LAM, 13013 Marseille, France}
\affil[c]{Leiden Observatory, Leiden University, Einsteinweg 55, Leiden, The Netherlands}
\affil[d]{University of Arizona, Steward Observatory, 933 North Cherry Avenue,Tucson, AZ 85719, USA}
\affil[e]{National Astronomical Observatory of Japan, National Institutes of Natural Sciences, Subaru Telescope, 650 North A’oh-ok-u Place, Hilo, HI 96720, U.S.A}
\affil[f]{W. M. Keck Observatory, Hawaii, U.S.A.}
\affil[g]{Jet Propulsion Laboratory, California Institute of Technology, 4800 Oak Grove Dr, Pasadena, CA 91109}

\authorinfo{Further author information: \\Vincent Chambouleyron: vchambou@ucsc.edu\\}

\begin{document} 
\maketitle

\begin{abstract}
The Zernike wavefront sensor (ZWFS) stands out as one of the most sensitive optical systems for measuring the phase of an incoming wavefront, reaching photon efficiencies close to the fundamental limit. This quality, combined with the fact that it can easily measure phase discontinuities, has led to its widespread adoption in various wavefront control applications, both on the ground but also for future space-based instruments. Despite its advantages, the ZWFS faces a significant challenge due to its extremely limited dynamic range, making it particularly challenging for ground-based operations. To address this limitation, one approach is to use the ZWFS after a general adaptive optics (AO) system; however, even in this scenario, the dynamic range remains a concern.
This paper investigates two optical configurations of the ZWFS: the conventional setup and its phase-shifted counterpart, which generates two distinct images of the telescope pupil. We assess the performance of various reconstruction techniques for both configurations, spanning from traditional linear reconstructors to gradient-descent-based methods. The evaluation encompasses simulations and experimental tests conducted on the Santa cruz Extreme Adaptive optics Lab (SEAL) bench at UCSC. Our findings demonstrate that certain innovative reconstruction techniques introduced in this study significantly enhance the dynamic range of the ZWFS, particularly when utilizing the phase-shifted version.
\end{abstract}

\keywords{adaptive optics, Zernike wavefront sensor, wavefront reconstruction}

\section{Introduction}

The exploration of exoplanets has become a key research topic in modern astronomy. Among the numerous techniques employed for exoplanet detection and characterization, direct imaging has emerged as a promising avenue, offering insights into the atmospheric composition and features, and potential habitability of distant worlds \cite{GPI,sphere}. Over the past few decades, significant improvements have been made to directly image exoplanets, primarily through advancements in instrumental technologies. Central to the success of direct imaging is the utilization of high-contrast imaging instruments, which enable the observation of faint planetary signals against the bright glare of their host stars. These instruments rely on precise wavefront sensing and control techniques and a device called coronagraph to suppress the starlight and reveal the much fainter planetary companions.

Among all wavefront sensors (WFSs), the Zernike wavefront sensor (ZWFS) stands out for it exceptional sensitivity, making it extremely robust to noise in the measurements \cite{ZeldaMamadou}. It also possesses the  ability to detect a wide range of aberrations, particularly phase discontinuities, a property not commonly found in other WFSs. These advantages makes the ZWFS an extremely good candidate for playing a key part in the next generation of high-contrast instruments, whether ground-based or space-based. One the main drawback of this sensor is its restricted dynamic range. This limitation arises from two fundamental aspects of the Zernike wavefront sensor. Firstly, the ZWFS is efficient in creating a proper reference wave only when aberrations in the wavefront are small. This self-reference mechanism, integral to the ZWFS operation, becomes less effective in scenarios where the wavefront exhibits significant deviations from the nominal state. Secondly, its interferometric nature results in a response that can be described by a sine wave, which is invertible only on a small interval. Detailed explanation of the ZWFS measurement principle will be provided in the next section.

In this paper, we propose a novel approach to enhance the dynamic range of the ZWFS for high-contrast imaging of exoplanets. Our method involves the implementation of a phase-shifted ZWFS (PSZWFS) configuration, wherein two ZWFS units are employed with slightly different settings. By introducing diversity into the wavefront sensing process, we aim to broaden the range of aberrations that can be accurately measured, thereby extending the dynamic range of the sensor. In the section \ref{section2}, we describe the principle of the phase-shifted ZWFS and derives basic equations and conditions to make it work, we also explain how such a sensor can be produced by using a vector-ZWFS, first proposed by \cite{Doelman:19}. Section \ref{section3} is presenting two efficient ways to reconstruct the signals of the phase-shifted ZWFS. In section \ref{section4}, we use simulations to demonstrate the expected gain in dynamic or our setup. Finally, we show some experimental results obtained on the SEAL testbed in section \ref{section5}.

\section{The phase-shifted Zernike WFS}
\label{section2}

\subsection{Equations for the Zernike WFS}

Before introducing the PSZWFS, this subsection aims at recalling the principle of the ZWFS measurements and detailing basic equations already derived in previous study \cite{ZeldaMamadou}. Here, we define the ZWFS class with a slightly more general definition: these are the FFWFS for which the complex-amplitude of the mask is composed of a central patch that can take any complex values (provided that the amplitude is not greater than 1) in an area of few $\lambda/D$ (see figure \ref{fig:layout}) and that is equals to 1 outside. We use this definition proposed in \cite{Haffert_psi} as it can encapsulates more complex ZWFS mask shapes as the ones proposed in \cite{rico_OWFS,chambou_SPIE2022}.

\begin{figure}[h!]
\centering
        \includegraphics[width=1\columnwidth]{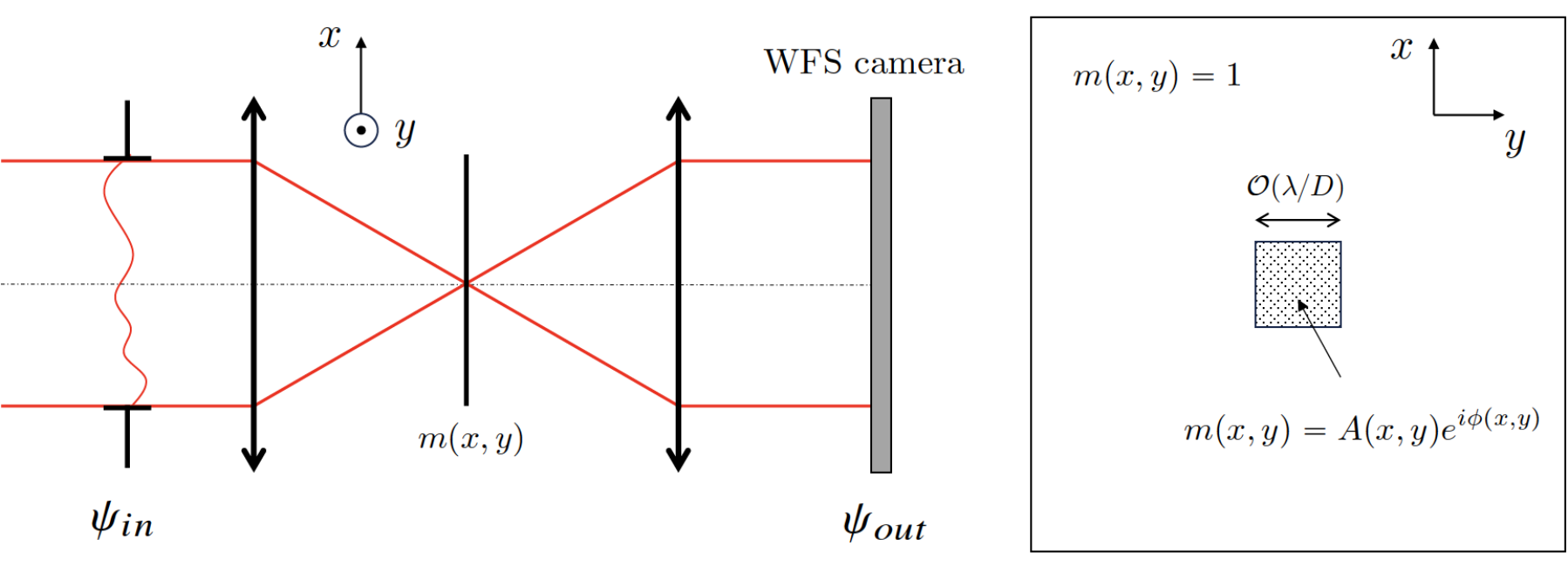}
    \caption{ZWFS basic optical layout. We use here a general definition of the ZWFS: a mask with a central patch (dashed-area on right figure) is placed in an intermediate focal plane to produce Fourier-filtering operation on the incoming EM field.}
    \label{fig:layout}
\end{figure}

Derivations are taken from \cite{Haffert_psi}. Let's write $\psi_{out}$ the electromagnetic (EM) field at the location of the WFS camera. We can write:

\begin{equation}
    \psi_{out} = \mathcal{F}^{-1}[\mathcal{F}[\psi_{in}]\times m]
    \label{eq:ZWFS_propag}
\end{equation}

where $\psi_{in}$ is the incoming EM field, $\mathcal{F}$ is the Fourier transform, $\mathcal{F}^{-1}$ the inverse Fourier transform and $m$ the complex-amplitude of the mask. By simply writing $m$ as $m = (m-1)+1$, we can modify the previous equation such as:

\begin{equation}
    \psi_{out} = \mathcal{F}^{-1}[\mathcal{F}[\psi_{in}]\times (m-1)] + \psi_{in}
\end{equation}

and we can then write 

\begin{equation}
    \psi_{out} = \psi_{ref} + \psi_{in}
\end{equation}

where:

\begin{equation}
    \psi_{ref} = \mathcal{F}^{-1}[\mathcal{F}[\psi_{in}]\times (m-1)] 
\end{equation}

And by defining the corresponding intensities of the different quantities as their square modulus ($I_{out} = |\psi_{out}|^{2}$, $ I_{ref} = |\psi_{ref}|^{2}$ and $I_{in} = |\psi_{in}|^{2}$), one can write the intensities recorded by the WFS camera by the general two-waves interference formula:

\begin{equation}
    I_{out} = I_{ref} + I_{in} + 2\sqrt{I_{ref}I_{in}}\cos\left(\phi_{in}-\phi_{ref}\right)
    \label{eq:ZWFS_response}
\end{equation}

And then, assuming that we have access to the quantities $I_{ref}$ and $\phi_{ref}$ (i.e we have the full knowledge of the reference EM field), it is possible to reconstruct the phase through the following equation:

\begin{equation}
    \phi_{in} = \phi_{ref} + \arccos \left(\frac{ I_{out}-I_{ref} - I_{in}}{2\sqrt{I_{ref}I_{in}}}\right)
\end{equation}

Although all the previous equations can apply to any FFWFS by making the central patch arbitrary large, using this last equation to perform wavefront reconstruction is useful only if the reference wave parameters are only slightly dependent on the incoming aberrations. Hence, this analysis makes sens only for masks having a central patch that extends at maximum few $\lambda/D$. It is therefore really suited for ZWFS-type reconstruction. In that case, a common approach is to assume that $\phi_{ref}$ corresponds to the one for a flat wavefront and $I_{ref}$ just being scaled by the square-root of the Strehl Ratio (SR). More sophisticated methods have proposed to update $I_{ref}$ and $\phi_{ref}$ by using an iterative algorithm \cite{Steeves:20}, and we will come back on that later in section \ref{section:iterative}.

\subsection{Principle of the phase-shifted Zernike WFS}
\label{section:PSZWFS}

One of the main limitation of equation \ref{eq:ZWFS_response}, is the that the function cosine is invertible only on a interval of length $\pi$. In order to break this degeneracy and extend the dynamic range, it is necessary to inject some diversity in the signal. A solution recently proposed is to inject know phases on top of the phase-to-be-measured \cite{Haffert_psi}. In this paper, we propose a more straightforward technique, consisting in injecting this phase diversity by using simultaneously two different ZWFS masks to perform the wavefront measurement.\\

Let's assume that we are splitting light of the incoming wavefront on two ZWFS mask configurations: configuration $m^{a}$ and configuration $m^{b}$ (we will present in the next paragraph an existing way to do that easily in practice). We can write the intensities recorded for each mask configuration:

\begin{equation}
\begin{split}
    I_{out}^{a} &= I_{ref}^{a} + I_{in} + 2\sqrt{I_{ref}^{a}I_{in}}\cos\left(\phi_{in}-\phi_{ref}^{a}\right)\\
        I_{out}^{b} &= I_{ref}^{b} + I_{in} + 2\sqrt{I_{ref}^{b}I_{in}}\cos\left(\phi_{in}-\phi_{ref}^{b}\right)
\end{split}
\end{equation}

Using a basic trigonometric identity, it is possible to write:

\begin{equation}
\begin{split}
\cos\left(\phi_{in}-\phi_{ref}^{a}\right) &= \cos(\phi_{ref}^{a})\cos(\phi_{in})+\sin(\phi_{ref}^{a})\sin(\phi_{in})\\
\cos\left(\phi_{in}-\phi_{ref}^{b}\right) &= \cos(\phi_{ref}^{b})\cos(\phi_{in})+\sin(\phi_{ref}^{b})\sin(\phi_{in})
\end{split}
\end{equation}

Assuming we have an accurate estimation of $I_{ref}$ and $\phi_{ref}$ for each mask, the phase can be estimated on its full range $[0,2\pi]$ by solving for each pixels in the measurements the following system of equations (this set of equations is really similar to the approach proposed by \cite{Haffert_psi}):

\begin{equation}
\begin{bmatrix}
\frac{ I_{out}^{a}-I_{ref}^{a} - I_{in}}{2\sqrt{I_{ref}^{a}I_{in}}}\\
\frac{ I_{out}^{b}-I_{ref}^{b} - I_{in}}{2\sqrt{I_{ref}^{b}I_{in}}}
\end{bmatrix}
= 
\begin{bmatrix}
 \cos(\phi_{ref}^{a}) &  \sin(\phi_{ref}^{a})\\
\cos(\phi_{ref}^{b}) &  \sin(\phi_{ref}^{b})
\end{bmatrix}
\begin{bmatrix}
\cos (\phi_{in})\\
\sin (\phi_{in})
\end{bmatrix}
=
M
\begin{bmatrix}
\cos (\phi_{in})\\
\sin (\phi_{in})
\end{bmatrix}
\label{eq:Phase-Shifted}
\end{equation}

The matrix $M$ determinant can be computed:

\begin{equation}
det_{M} = 
\begin{vmatrix}
 \cos(\phi_{ref}^{a}) &  \sin(\phi_{ref}^{a})\\
\cos(\phi_{ref}^{b}) &  \sin(\phi_{ref}^{b})
\end{vmatrix}
=  \sin(\phi_{ref}^{b}-\phi_{ref}^{a})
\label{eq:det}
\end{equation}

The matrix is therefore not invertible for positions for which $\phi_{ref}^{b}\equiv \phi_{ref}^{a} \mod \pi$. And maximum value for the determinant for the positions where: $\phi_{ref}^{b}\equiv \phi_{ref}^{a} \mod \pi/2$.\\

Let's apply these results to a classic ZWFS configuration where the central patch is a circular dimple of constant phase and amplitude 1. Then, $m$ can be written as: $m=\mathcal{H}e^{i\theta}+(1-\mathcal{H})$ where $\mathcal{H}$ is the Heavyside function (describing the dimple). Then we have:

\begin{equation}
    \psi_{ref} = (e^{i\theta}-1)\cdot \mathcal{F}^{-1}[\mathcal{F}[\mathcal{H}]]
\end{equation}

Using the identity:
\begin{equation}
e^{i\theta}-1 = 2\cdot i \sin \left(\frac{\theta}{2}\right)e^{i\theta/2}
\end{equation}

we can re-write the reference beam as:

\begin{equation}
    \psi_{ref} = 2\cdot i \sin \left(\frac{\theta}{2}\right)e^{i\theta/2}\cdot \mathcal{F}^{-1}[\mathcal{F}[\mathcal{H}]]
\end{equation}

Therefore, we see that if we have two masks with same dimple size and phase shifts being $\theta_{a}$ for one and $\theta_{b}$ for the other, the difference in phase for the reference beams produced will be $\Delta \phi_{ref} = (\theta_{b}-\theta_{a})/2 = \Delta \theta /2 $ for each positions of the measurements. Therefore if we consider two ZWFS mask configurations that have WFS capabilities, the phase-shifting technique will work if phase shift difference is different from $\Delta \theta = 0 \mod 2\pi$ (which means that the two masks are the same, hence not producing phase diversity) and will be optimal for phase-shift difference of $\Delta \theta = \pi \mod 2\pi$ between the two masks, for which $det_{M}$ is maximal (equation \ref{eq:det}).\\

Thus, we have shown that using two different ZWFS mask configuration could help in theory to increase dynamic range. The two different masks could have radically different shapes, but a straightforward configuration is to use classic ZWFS masks of same dimple size while using a differential phase-shift of their central dimple of $\Delta \theta = \pi \mod 2\pi$. 

\subsection{A practical implementation: the vector-Zernike WFS}
\label{section:vZWFS}

A practical implementation of such a phase-shifted ZWFS is the vector ZWFS (vZWFS), first proposed by \cite{Doelman:19}. The concept rely on using the same mask for both ZWFS configurations by manufacturing a mask that induced different phase-shift according to incoming polarization states. To our knowledge, there are mainly two ways in fabricating such devices: either using liquid crystals using metasurfaces masks. \\

\textbf{Liquid crystals:} the first proposition of the vZWFS relied on this technology, consisting in impregnating different phase shifts according to circular polarization. Circular polarizations are then transformed in orthoganl one by going through a quarter-wave plate and then are splitted by a wollaston prism (see figure \ref{fig:layout}). In this paper, it is this kind of technology used for the experimental demonstrations on the SEAL testbed and proposed in \cite{Doelman:19} when vZWFS was first introduced.\\

\textbf{Metasurfaces:} another way to build a vZWFS is to use a metasurfaces mask, composed of sub-wavelength structures that will induce a different phase shift according to the orthogonal polarizations directly \cite{metaZWFS}. This polarization are then split by a wollaston prism (see figure \ref{fig:layout}).

\begin{figure}[h!]
\centering
        \includegraphics[width=0.8\columnwidth]{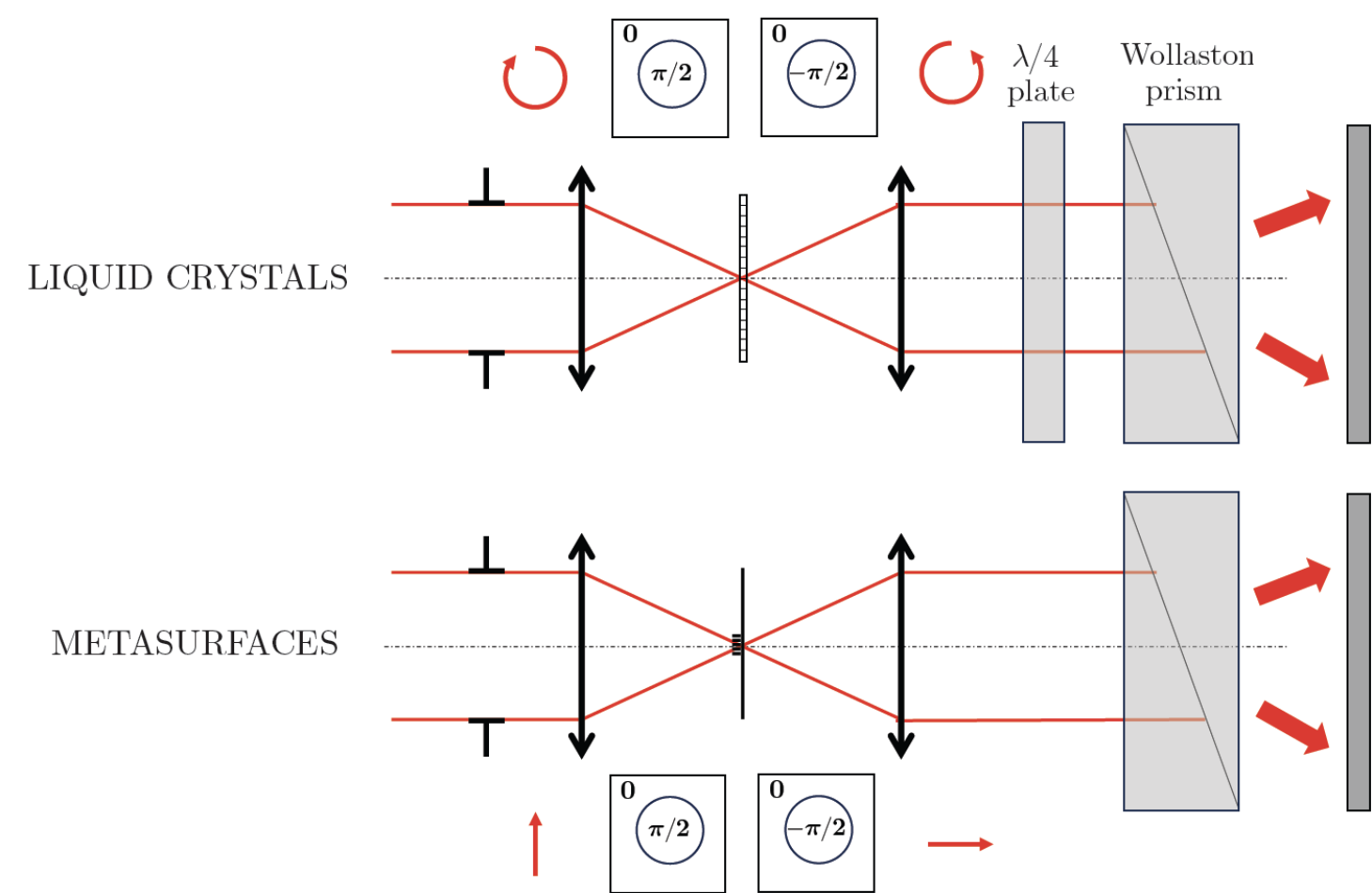}
    \caption{Implementing the PSZWFS using the vector-ZWFS approach. \textbf{Top:} Liquid crystal approach, as used in \cite{Doelman:19}. \textbf{Bottom:} Metasurfaces approach, as implemented by \cite{metaZWFS}.}
    \label{fig:layout}
\end{figure}

It is worth noting that noth techniques encounter a common challenge related to the use of a wollaston prism: when attempting to image both pupils on the same detector, a slight focus shift occurs between the two images.

\section{Reconstruction methods proposed}
\label{section3}

All reconstruction methods proposed here have a common ground: they use a digital twin of the PSZWFS, that is to say a high-fidelity numerical model of the wavefront sensor.

\subsection{Iterative arcsine phase-shifted approach}
\label{section:iterative}
This method is based on equation \ref{eq:Phase-Shifted}. Assuming that we have access to the shape of the reference waves coming from each ZWFS, inverting matrix $M$ for each pixel can gives us access to an estimation of the $\cos(\phi_{in})$ and $\sin(\phi_{in})$. The idea is to perform an iterative approach: at each step the previous estimation of $\phi_{in}$ is injected in the numerical models of the ZWFS to create a new estimation of their respective references waves and re-evaluate the phase through equation equation \ref{eq:Phase-Shifted}. 

\begin{enumerate}
    \item We first assume a flat wavefront to estimate the reference wave $I_{ref}$ and $\phi_{ref}$ for both configurations ZWFS$^{a}$ and ZWFS$^{b}$.
    \item For each pixels in the pupil (in the ZWFS detector plane), solving the set of equations \ref{eq:Phase-Shifted} to find an estimate of the phase.
    \item Injecting the estimated phase to update reference waves for both configurations ZWFS$^{a}$ and ZWFS$^{b}$.
    \item Go back to step 2 and iterate.
\end{enumerate}

\subsection{Gerchberg-Saxton algorithm}

The GS algorithm was first proposed by \cite{gerchberg1972practical} and is widely used to perform image sharpening from point spread function (PSF) images \cite{Fienup:82,keck_GS}, but was also recently applied as a non-linear reconstructor for the pyramid WFS \cite{chambou_GS}. The principle of the GS algorithm is to propagate the light back and forth in the numerical models of the two ZWFS configurations, while injecting the knowledge of the amplitudes (entrance pupil and measurement plane) of the complex quantities we are trying to retrieve at each step. The relation between the complex amplitude in the input pupil plane and the measurement plane is given by equation \ref{eq:ZWFS_propag}.

One iteration of the GS algorithm can be split into four steps:

\begin{enumerate}
    \item For each ZWFS configuration: we compute the EM-field amplitude $A_{out}$ by simply applying $\sqrt{I_{out}}$ to the ZWFS measurements. For the first iteration, the complex EM-field in the detector plane is built by using as a starting phase $\phi_{out}=\arg(A_{in} \star \widehat{m})$, which corresponds to the phase in the detector plane when a flat wavefront is propagated through the ZWFS system. Therefore, we have a first estimate $\psi_{out}$ that can be back-propagated in the system through the following equation (corresponding to the back propagation of equation \ref{eq:ZWFS_propag}):

\begin{equation}
    \psi_{in} = \mathcal{F}^{-1}[\mathcal{F}[\psi_{out}]\times \bar{m}]
    \label{eq:ZWFS_propag}
\end{equation}

where $\bar{\cdot}$ is the complex conjugate operator.

    \item A first estimation of the EM-field $\psi_{in}^{a}$ (resp. $\psi_{in}^{b}$) is obtained for the ZWFS$^{a}$ (resp. ZWFS$^{b}$) configuration. The total EM-field is then computed as: 
    
    \begin{equation}
    \psi_{in} = \frac{1}{\sqrt{2}} (\psi_{in}^{a}+\psi_{in}^{b})
    \label{eq:Back_propag}
    \end{equation}
 
    \item Because we already have access to $A_{in}$ , the amplitude found through back-propagation is discarded and replaced by the measurement of $A_{in}$ while keeping the estimated phase $\phi_{in} = \arg (\psi_{in})$. The entrance pupil plane EM-field can then be propagated in the system (direct propagation, through equation \ref{eq:ZWFS_propag}).
    \item A new estimation of $\psi_{out}$ is obtained. As previously done for the entrance pupil plane, we discard the amplitude and replace it by the measurement of $A_{out}$ given by the detector and keep the estimated phase $\phi_{out}$. We can then return to step 1 and iterate again. We call one iteration of the GS algorithm the numerical operation that consists of these four steps. 
\end{enumerate}

\subsection{Gradient descent approach}

The method is really general and could in principle work with any optical system. It consists in running a gradient descent that optimize input phase so the simulated measurements match the observed ones. For that, we choose as cost function $\mathcal{C}$ the mean square error between the true measurements and the simulated one:

    \begin{equation}
    \mathcal{C} = ||D-I||_{2}
    \label{eq:Back_propag}
    \end{equation}

where $||\cdot||_{2}$ is the 2-norm, $D$ is the true measurements and $I$ the simulated ones. To drastically boost convergence speed, we use the method proposed in \cite{Haffert_nonlinear} consisting in computing the gradient of the cost function by applying back-propagation through the ZWFS model. The computed analytically computed gradient is then feed to an optimizer (in our case the \textit{scipy.minimize} function from python).

\section{Simulations}

\label{section4}

\subsection{Gains in dynamic range}
\label{section:dynamics}  

We first tested in simulation the different reconstructors presented in the previous section. This simulations are ideal as they don't exhibit any mismatch between the ZWFS configuration and its version used for the reconstruction. To demonstrate the improved dynamics offered by the PSZWFS and the proposed reconstructors, we considered two ZWFS configurations: (i) a classic ZWFS configuration, with only one pupil, with a mask of dimple diameter $d=1.06\lambda/D$ and $\theta = \pi/2$ (ii) a PSZWFS configuration with 2 masks of dimple diameter $d=1.06\lambda/D$  and phase-shifts of $\pi/2$ and $-\pi/2$. All iterative algorithms are performed with 25 iterations.

\begin{figure}[h!]
\centering
        \includegraphics[width=0.6\columnwidth]{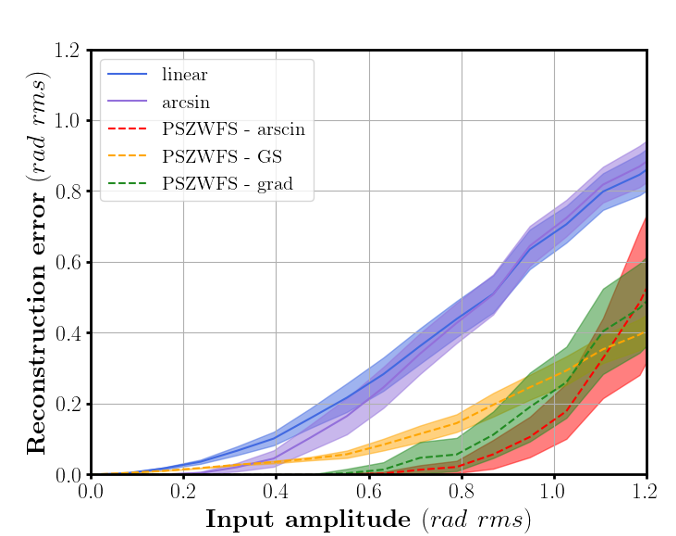}
    \caption{Linearities curves for the ZWFS and the PSZWFS in the case of different reconstructors.}
    \label{fig:linearity_simu}
\end{figure}

We simulated dynamic range plots by generating randomly sampled aberrations, based on a $f^{-2}$ power spectral density (PSD) defined on the first 200 Zernike polynomials (the measurement plane being over-sampled at 100 pixels across diameter to avoid any aliasing nor fitting error), that were then presented to the ZWFS systems, with a total wavefront error varying from 0 to 1.3 rad rms. In order to avoid impact of phase unwrapping in this analysis, which occurs around 1 radians rms depending on the phase realization, reconstructed phase was compared to the input wrapped phase. For each wavefront error value, 20 phase screens were injected and reconstructed. This process was carried out for the ZWFS with a linear reconstructor (interaction matrix approach) and an arcsine iterative reconstructor \cite{Steeves:20}, \cite{Haffert_psi}, and for a PSZWFS configuration, using the arcsine iterative phase-shifted approach, the GS algorithm and the gradient descent algorithm. The mean values of the 20 reconstructions for each input amplitude in all configurations are shown in Figure \ref{fig:linearity_simu}, with their associated standard deviation represented by the shaded areas. As shown in \cite{Haffert_psi} for the case of the classic ZWFS, the iterative arcsine reconstructor performs better than the linear reconstructor but still presents issues after $~0.3$ radians rms where degeneracy in the ZWFS signal impacts drastically the reconstruction. However, the PSZWFS and associated non-linear reconstruction increases drastically the dynamic range. One can notice that the GS algorithm presented a systematic offset in the reconstruction. This comes from the fact that part of the light in the measurement plane is diffracted beyond the detector spatial extension. Hence, back propagation doesn't take in account this light. The phase-shifted and gradient descent approaches perform similarly. If we arbitrary define a metric for satisfactory reconstruction as the point where the reconstruction error reaches 5\% of the input rms, the PSZWFS combined with non-linear algorithms extends this metric value by more than an order of magnitude compared to the ZWFS combined with a linear reconstructor.

It seems that gradient descent and arcsine iterative phase-shifted method exhibit similar performance. However, the first one is taking more time and can work only on a limited number of parameters in order to converge in a feasible amount of time. Overall the phase-shifted method is the more appealing one, but have one drawback: because its formula is derived from analytical equations, it is less straightforward to implement for a system slightly different (for example in the case where one of the pupils is out of focus), whereas the GS and gradient descent techniques can be applied to any optical system, provided the user has a good model available. One other advantage of the gradient descent method, not considered here, is the fact that it can perform both phase and amplitude retrieval \cite{Haffert_nonlinear}.

To conclude on this dynamics range study in simulation, it is clear that the PSZWFS combined with non-linear reconstruction techniques can drastically improved dynamic range of the classic ZWFS. The reconstructors presented here are not suited for real time are they require to perform several Fourier transforms to reconstruct the phase. Despite that, these reconstructors show that there is a way to push PSZWFS reconstruction up to about 1 radians rms. After this value, the aberrations are too important for the ZWFS to produce a useful reference beam to encode the input phase.  These kind of reconstructors could be approach by machine learning algorithms \cite{Landman:20,2022A&A...664A..71N,2023ML}, more suited for real time. In this simulation no model errors were introduced, which represents an idealized case for this kind of model-based reconstructors. We will extend more on model errors in the next section dedicated to experimental validation.

\subsection{PSZWFS variations}

One could wonder what kind of PSZWFS configurations works in terms of dimple characteristics. This answer was partially answered in section \ref{section:PSZWFS}.

\textbf{In terms of dimples phase-shifts :} As presented in section \ref{section:PSZWFS}, any masks that exhibits a differential phase shift different from $2\pi$ and that are not in the specific configuration $\theta_{a} = 0$ and $\theta_{b} = \pi$ (that would result in no intensities linear with the phase) works. However, the diversity signal increase when the phase difference approaches $\pi$ and some configuration of the ZWFS exhibits poor sensitivity \cite{chambou_Z2WFS}. These two aspects leads to the fact that the configuration being the more interesting in term of sensitivity is the one with the phase-shifts: $\pi/2$ and $-\pi/2$.

\begin{figure}[h!]
\centering
        \includegraphics[width=0.7\columnwidth]{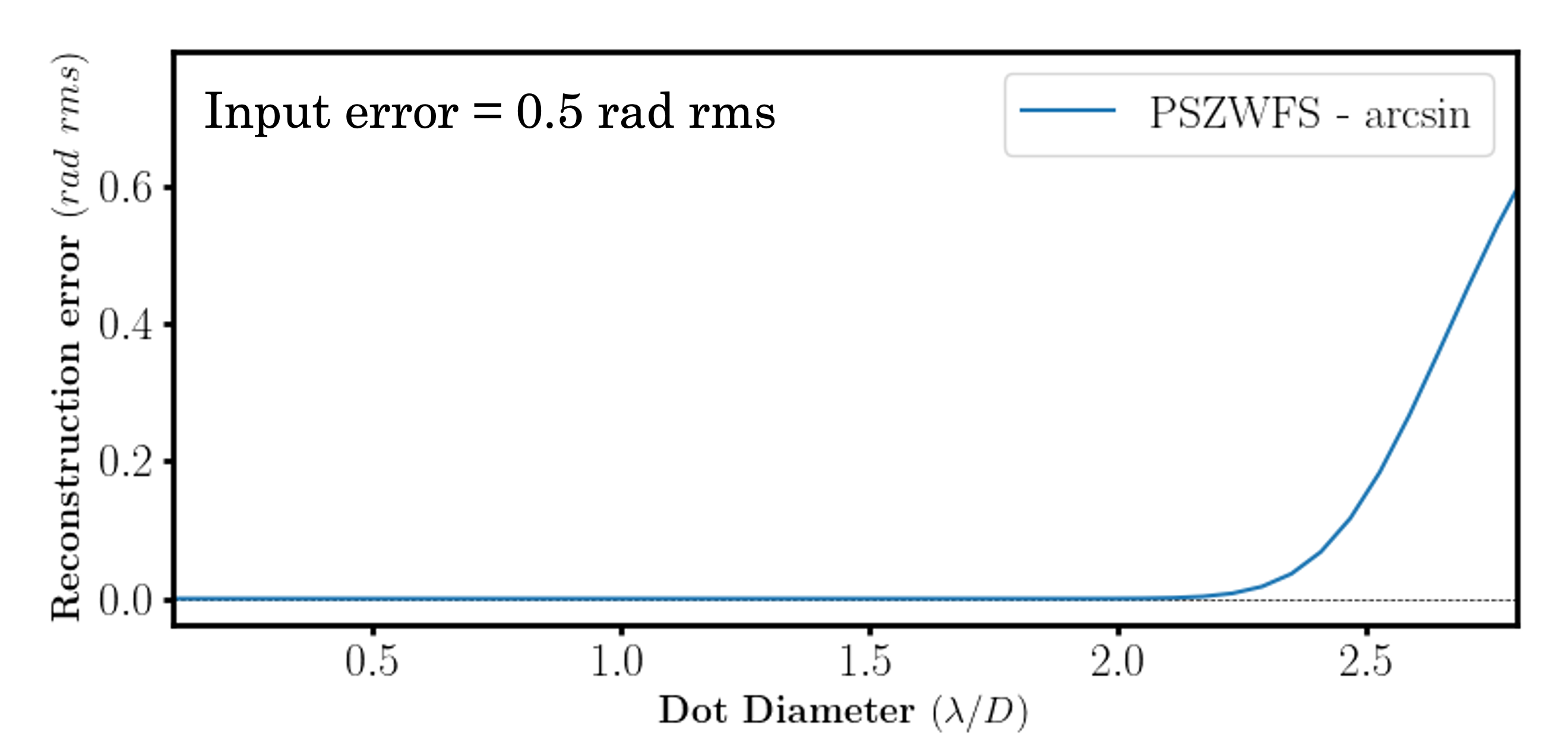}
    \caption{Impact of PSZWFS dimples diameter on phase reconstruction in the case of the iterative arcsine reconstructor.}
    \label{fig:diameter}
\end{figure}

\textbf{In terms of dimples diameter :} Studies have shown that the dimple diameter can impact sensitivity behavior with respect to spatial frequencies. Notably, increasing the diameter impact the capabilities to perform low-spatial frequencies sensing. We are here interested in how far the dimple diameter can be increased in the context of a PSZWFS configuration combined with the phase-shifted non-linear reconstructor. To perform this study in simulation, we used a input aberrations of 0.5 radians rms ($f^{-2}$ PSD) and perform the reconstruction. Results are given figure \ref{fig:diameter}, showing the reconstruction error versus increasing dimple diameter. We notice that after for the dimple diameter greater than about 2.2 $\lambda/D$ , the reconstruction error drastically increases. This is due to the fact that for too large dimple diameters, the sensitivity to low-order modes significantly decreases \cite{chambou_Z2WFS}. Hence, reconstruction for low-order mode is not accurate any more. However for this kind of non-linear iterative reconstruction which is using the knowledge of each mode to infer the next estimation, a mismatch of reconstruction for some modes will also affect the reconstruction accuracy for all other modes (it can be thought as modal confusion). Hence, we come to the conclusion that having a too large dimple is problematic for such reconstructors. We also precise that even if we only show results for the arcsine iterative method, the exact same behavior is observed for all the other kind of reconstructors considered in this study.

\section{Experimental validation}
\label{section5}

\subsection{Presentation of the SEAL testbed}

\begin{figure}[h!]
\centering
        \includegraphics[width=0.8\columnwidth]{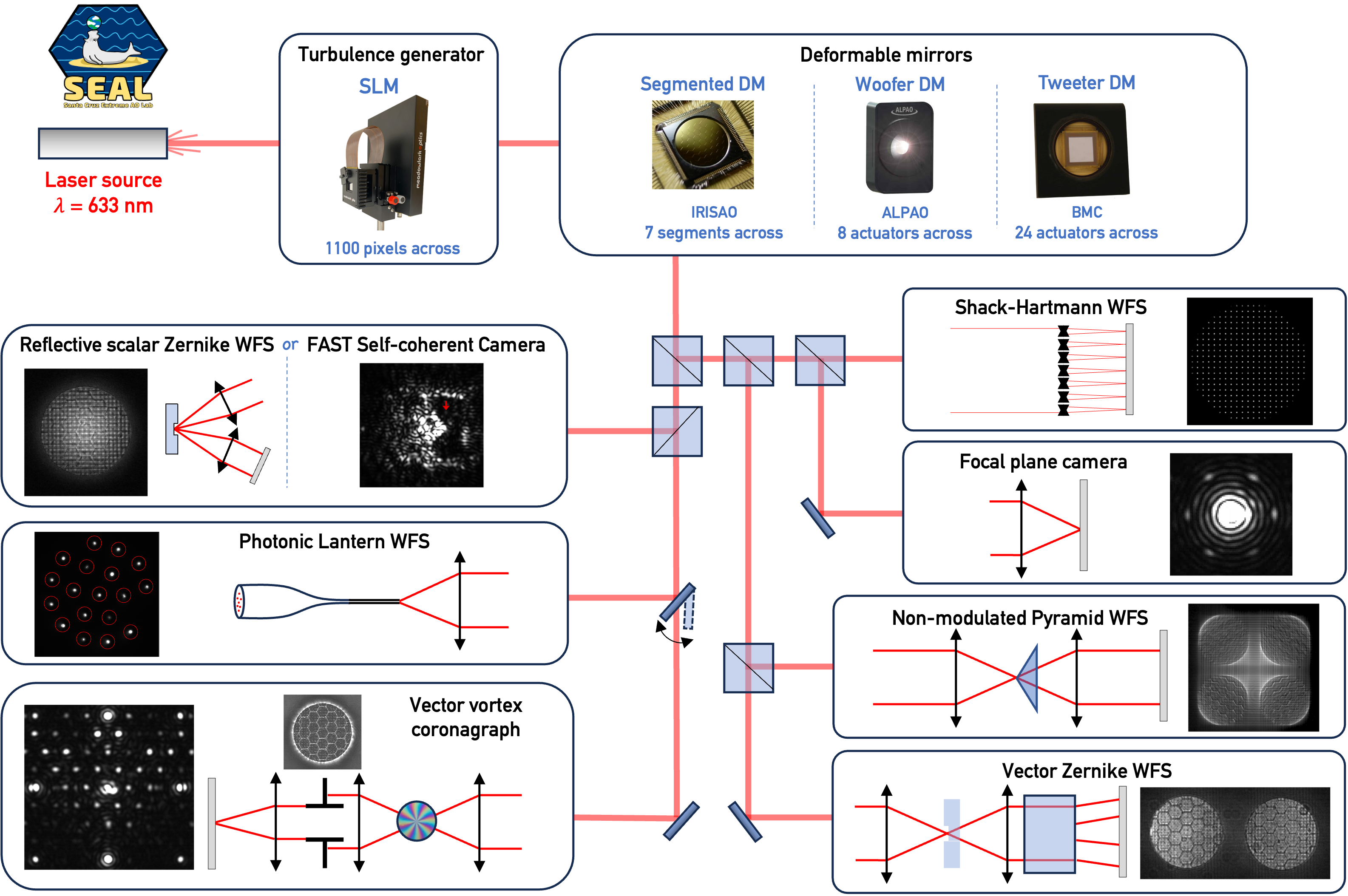}
    \caption{Simplified diagram of the SEAL testbed at UCSC.}
    \label{fig:SEAL}
\end{figure}

The objective of this section is to showcase a laboratory demonstration of the PSZWFS and the algorithms presented in the previous section. For that, we used the Santa Cruz Extreme Adaptive optics Laboratory (SEAL\cite{SEAL}), an advanced adaptive optics testbed equipped with multiple deformable mirrors (DM), wavefront sensors, and coronagraphic branches. The goal of this testbed is to test high-contrast wavefront sensing and control strategies for ground-based segmented telescopes. Figure \ref{fig:SEAL} depicts a simplified schematic layout of the SEAL testbed. Among all the SEAL components, the ones pertinent to this experiment include a light source at $\lambda = 635\ nm$, an irisAO segmented DM with six segments across the pupil diameter, and a vZWFS.

\begin{figure}[h!]
\centering
        \includegraphics[width=0.4\columnwidth]{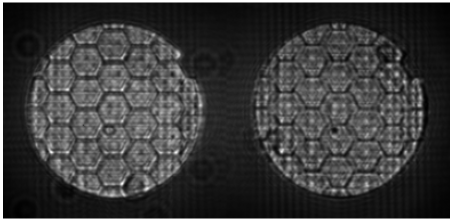}
    \caption{SEAL vZWFS reference intensities for the a flat wavefront.}
    \label{fig:ZWFS_I0}
\end{figure}

The PSZWFS is therefore implemented through a vZWFS mask (explanation of the principle was given section \ref{section:vZWFS}), the same used in \cite{Doelman:19}. Signals recorded by the vZWFS for a flat wavefront are given figure \ref{fig:ZWFS_I0}. One can notice the segment gaps from the irisAO DM, and the a high-order grid structure coming from the well-known "quilting" effect of the BMC DM. The pupil is made circular by a circular pupil plane stop located at the entrance of the testbed. It is also possible to notice that there is a missing segment on the top-right part of the pupil, due to a failure of this segment on the irisAO DM, inducing such a large tilt that light coming from this segment is not propagated through the optical system.

\subsection{Experimental results}

\begin{figure}[h!]
\centering
        \includegraphics[width=1\columnwidth]{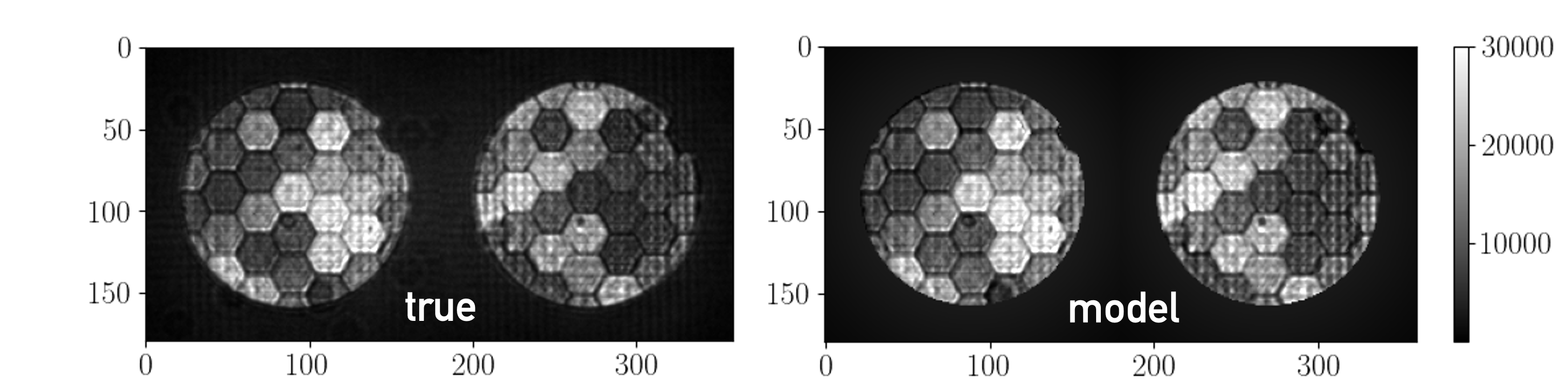}
    \caption{Comparison between true SEAL vZWFS measurements and simulated intensities through the vZWFS numerical model.}
    \label{fig:seal_signals}
\end{figure}

In order to experimentally test the non-linear reconstructors proposed in the previous section for the vZWFS installed on SEAL, we used the irisAO DM to send known phase aberrations on the WFS. The choice of the irisAO DM was motivated by the fact that this kind of DM are very well calibrated and allows to have a good control on the phase sent in the system (in other words, the true phase can be assumed to be the commands sent to the irisAO DM). The model of SEAL vZWFS is generated by recording an off-mask pupil image and simulating optical propagation in the sensor assuming Fraunhofer diffraction. In the presence of segment piston aberrations generated by the irisAO DM, figure \ref{fig:seal_signals} shows the comparison between the true image recorded on the bench and the simulated image, generated by propagating the reconstructed phase in the vZWFS model.

\begin{figure}[h!]
\centering
        \includegraphics[width=0.75\columnwidth]{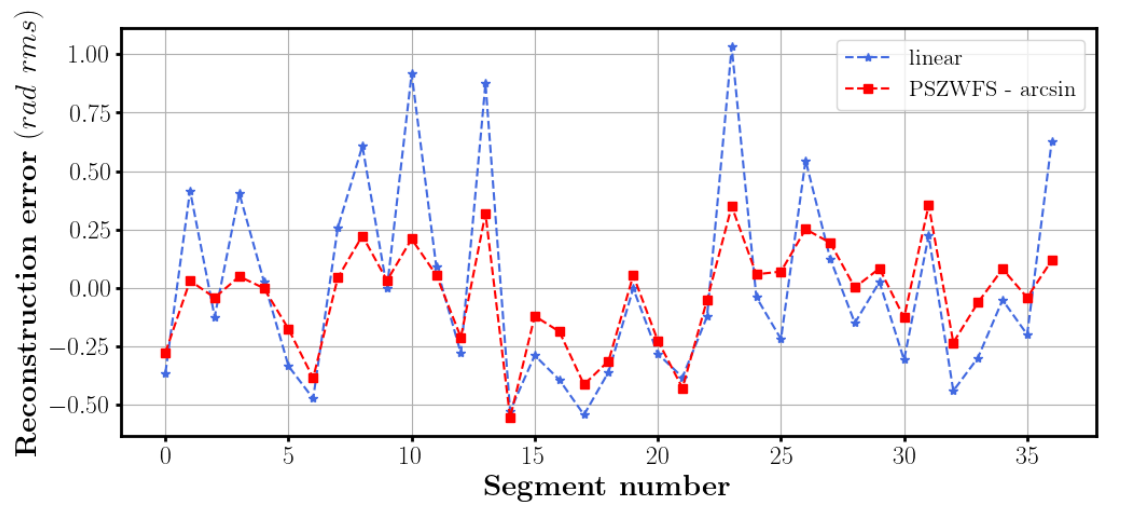}
    \caption{Reconstruction error for each of the SEAL 36 segments for the linear ZWFS reconstruction and for non-linear PSZWFS reconstruction.}
    \label{fig:segment_error}
\end{figure}

We reproduce the same procedure described in section \ref{section:dynamics} to assess dynamic range of the different reconstructor: aberrations following a $f^{-2}$ power-law were generated and then projected on irisAO DM segments using only segment pistons. Retrieved phase is projected on the irisAO DM segments and compared to input commands to estimate reconstruction errors. All measurements are taken for high signal-to-noise ratio, in order to assess only linearity behavior. We also generated simple ZWFS reconstruction by taking only one of the two vZWFS's pupils for the reconstruction. An example of phase reconstruction error comparison for one given input phase of $0.8$ radians rms for the linear reconstructor and the iterative arcsine for the PSZWFS is shown figure \ref{fig:segment_error}.

\begin{figure}[h!]
\centering
        \includegraphics[width=0.75\columnwidth]{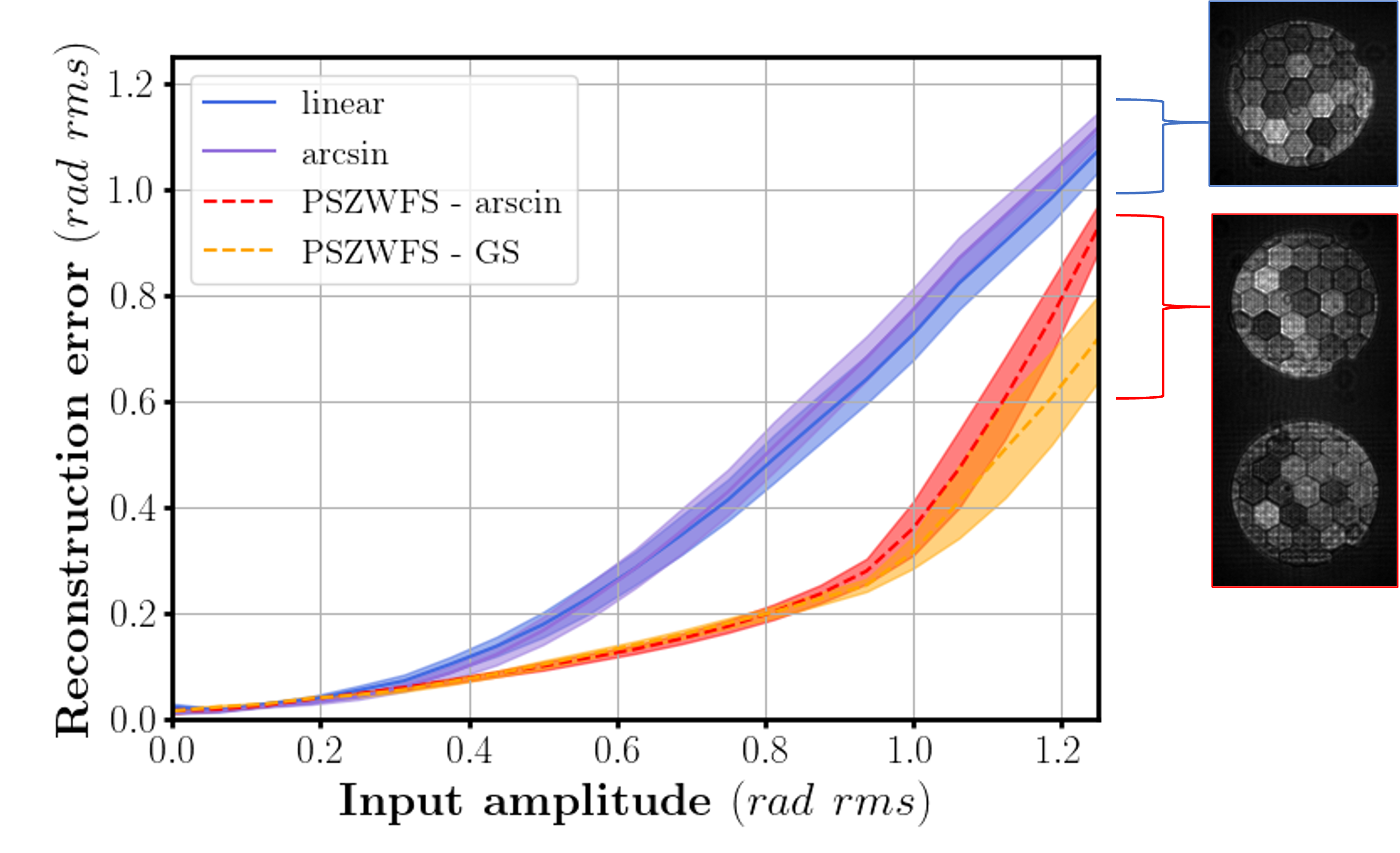}
    \caption{Experimental linearities curves for the ZWFS and the PSZWFS in the case of different reconstructors on SEAL. The classic ZWFS reconstructions correspond to the use of only one of the two vZWFS pupil.}
    \label{fig:linearity_SEAL}
\end{figure}

All reconstruction results are given figure \ref{fig:linearity_SEAL}, where we compare: (i) the linear and arcsine approaches for the simple ZWFS (ii) the GS and iterative arcsine approaches for the vZWFS (gradient descent method has not been experimentally implemented yet). We observe that the linear approach and the arcsine give similar performance for the classic ZWFS, while vZWFS methods are indeed increasing dynamic range. However, the model-based reconstructors don't perform as good as in simulation and seems to present systematic offsets in the regime where they should give near zero reconstruction error. That is also explaining why the arcsine reconstructor for the simple ZWFS doesn't perform better than the linear reconstructor, unlike simulations presented figure \ref{fig:linearity_simu}. This under-performance can be explained by model errors that could originates from several parameters. First, there is some polarization cross-talks between the two vZWFS pupils. As a matter of fact, a small misalignment between the mask the quarter-wave plate could lead to a leakage of one pupil signal into another (see figure \ref{fig:layout}). This small leakage can lead to an underestimation of the phase as part of the signal could be canceled out because of this cross-talk. The slight defocus on the vZWFS pupil images are also not taken in account in our model. Finally other models errors linked to the optical layout and also the detector could impact the performance.

The next step for this experimental work is logically to understand the exact sources of our model error in order to improve the reconstruction algorithms to approach simulations. It is also planned to implement gradient descent, not only for phase reconstruction but also to numerical fit our model parameters to better match the true SEAL vZWFS. 

\section{Conclusion}

The ZWFS is one of the most sensitive wavefront sensors available for high-contrast imaging but is limited by a narrow dynamic range. This limited range arises from the interferometric nature of the ZWFS, leading to signal degeneracies when measuring phases with larger amplitudes. We demonstrated that employing two ZWFS units in parallel, each with different phase shifts, introduces phase diversity into the measurements, thereby breaking these degeneracies. This approach called the phase-shifted ZWFS (PSZWFS), allows for the accurate reconstruction of larger phase amplitudes. The PSZWFS can be readily implemented using vZWFS configuration, either by using liquid crystals or metasurfaces.

Our simulations showed that combining the PSZWFS with nonlinear reconstruction methods significantly enhances the dynamic range of the classic ZWFS. This improvement was also demonstrated experimentally on the SEAL testbed, although the performance was somewhat below the simulation predictions due to model inaccuracies.

The next steps in this research are twofold: firstly, refining the numerical model of the vZWFS to enhance the accuracy of SEAL reconstructions; and secondly, addressing the main limitation of the current reconstruction methods—their unsuitability for real-time application. The iterative algorithms used here are effective for calibration and controlling slowly evolving aberrations but are not yet fast enough to operate at the kilohertz speeds required for second-stage AO applications. To overcome this, one promising approach is to approximate these nonlinear reconstructors with machine learning models, which can significantly reduce inference times and enable real-time operation.

\acknowledgments 
 
This work benefited from the 2024 Exoplanet Summer Program in the Other Worlds Laboratory (OWL) at the University of California, Santa Cruz, a program funded by the Heising-Simons Foundation and NASA. 

\bibliography{report} 
\bibliographystyle{spiebib} 

\end{document}